# Multiplexed sub-Doppler spectroscopy with an optical frequency comb

D. A. Long,[1,*] A. J. Fleisher,[1,*] D. F. Plusquellic,[2] and J. T. Hodges[1]

[1]*Material Measurement Laboratory, National Institute of Standards and Technology, 100 Bureau Drive, Gaithersburg, Maryland 20899, USA*
[2]*Physical Measurement Laboratory, National Institute of Standards and Technology, 325 Broadway, Boulder, CO 80305, USA*

*Corresponding authors: David A. Long (david.long@nist.gov; Tel.:(301)-975-3298; Fax: (301)-869-4020) and Adam J. Fleisher (adam.fleisher@nist.gov; Tel.:(301)-975-4864; Fax: (301)-869-4020).

## Abstract

An optical frequency comb generated with an electro-optic phase modulator and a chirped radiofrequency waveform is used to perform saturation and pump-probe spectroscopy on the $D_1$ and $D_2$ transitions of atomic potassium. With a comb tooth spacing of 200 kHz and an optical bandwidth of 2 GHz the hyperfine transitions can be simultaneously observed. Interferograms are recorded in as little as 5 µs (a timescale corresponding to the inverse of the comb tooth spacing). Importantly, the sub-Doppler features can be measured as long as the laser carrier frequency lies within the Doppler profile, thus removing the need for slow scanning or *a priori* knowledge of the frequencies of the sub-Doppler features. Sub-Doppler optical frequency comb spectroscopy has the potential to dramatically reduce acquisition times and allow for rapid and accurate assignment of complex molecular and atomic spectra which are presently intractable.

Sub-Doppler spectroscopy has been shown to be a powerful tool for optical frequency metrology [1-3], observing fundamental quantum light-matter interactions [4], assigning complex and blended spectra [5, 6], and laser frequency stabilization [7-9]. While reducing the widths of spectral features from their broad Doppler profiles down to or below their natural linewidths is immensely powerful, the ultra-narrow widths of the resulting features often result in long acquisition times where broad spectral regions must be slowly scanned with small frequency steps. The process becomes even more challenging when the nominal transition frequencies are not known *a priori*, as in the case of blended or multicomponent spectra.

A multiplexed approach for the acquisition of sub-Doppler spectra could lead to many-orders-of-magnitude reduction in acquisition time as well as allowing for applications to presently intractable problems such as sub-Doppler spectroscopy of the dense spectra of hydrocarbons or other large non-rigid molecules. While mode-locked optical frequency combs present a powerful platform for multiplexed Doppler-broadened spectroscopy [10, 11], their wide comb spacing (generally between 100 MHz and 1 GHz) is far too large to interrogate sub-Doppler features without a slow dither of the repetition rate [12]. More recently an external phase modulator has been used to reduce the comb mode spacing of a mode-locked laser comb [13]. Here we present an alternate approach in which a narrowly spaced optical frequency comb is generated through the use of an electro-optic phase modulator (EOM) [14-16] driven by a frequency chirped waveform from an arbitrary waveform generator (AWG). This frequency comb is used in a self-heterodyne configuration [17, 18] to simultaneously record saturation and optical pumping hyperfine transitions of $^{39}K$ with interferograms as short as 5 μs.

A schematic of the sub-Doppler direct frequency comb spectrometer can be found in Fig. 1. The laser source was an external-cavity diode laser with a linewidth less than 100 kHz (at 1 ms) and a tuning range of 745 nm to 785 nm. The laser power was 6 mW following injection into the polarization-maintaining fiber and subsequent optical isolators. The laser frequency was actively stabilized through the use of a high precision wavelength meter which provided a maximum resolution of 0.5 MHz and a data acquisition rate up to of 400 Hz. A software-based proportional-integral-derivative servo was used to lock the laser frequency to a given set point by feeding back to the laser current and piezoelectric transducer with a bandwidth of 100 Hz.

The optical frequency comb was generated through the use of a waveguide-based electro-optic phase modulator which was driven by an AWG with a sample rate of 10 GSamples/s. The electro-optic modulator had a response at greater than 20 GHz and $V_\pi$ = 2.6 V at 1 GHz. A train of linearly chirped waveforms was generated by the AWG which covered frequencies between 0.2 MHz and 1,000 MHz in 5 μs. Each waveform is given as a function of time by: $h(t) = A\sin\left\{2\pi\left(f_0 t + \frac{(f_1 - f_0)t^2}{2\Delta t} + \phi\right)\right\}$, where $A$ is a constant amplitude, $f_0$ and $f_1$ are the initial and final frequencies of the chirp, respectively, $\Delta t$ is the chirp duration, and $\phi$ is a constant phase term. This led to a nearly power-leveled optical frequency comb with a spacing of 200 kHz which spanned 2 GHz in bandwidth. We note that the generation of such a narrowly spaced comb using a corresponding mode-locked laser would require a laser cavity on the order of 0.7 km in length. Further, the use of chirped waveforms gives rise to interferograms with nearly constant amplitude in the time domain, removing the need for extremely wide dynamic range when using mode-locked laser combs (due to the large center burst) as well as reduced quantum shot noise [19]. Importantly, chirped waveforms such as these (or alternatively pseudo-random bit sequences [17, 18]) can be generated without an expensive AWG using low-cost direct digital synthesis [20] or segmented approaches [21].

A free-space acousto-optic modulator (AOM) was used to shift the carrier frequency of the resulting self-heterodyne signal away from DC by utilizing the first-order output at 384.22 MHz. In order to facilitate coherent time-domain averaging of the self-heterodyne signal, this carrier frequency was phase-locked to an external frequency reference by actuating the voltage-controlled oscillator which provided the AOM radiofrequency drive signal [22]. This lock was performed through the use of a phase frequency detector whose output was then fed into a commercial loop filter (proportional-integral corner at 1 kHz). This approach has been previously shown to enable more than two hours of coherent, real-time averaging, thus allowing for far more efficient data management [22].

The free-space laser beam was then double-passed through a 46 cm Brewster-angled (for the incoming linearly-polarized beam) sealed glass cell containing atomic potassium metal at natural isotopic abundance (99.5 % purity). At a temperature of 296 K, this gives rise to a vapor pressure of 1.5 μPa [23]. Approximately 0.09 mW to 0.15 mW of total optical power (i.e., the sum of all of the comb teeth and the carrier) was injected into the cell, corresponding to a laser intensity of 14 $mW/cm^2$ to 27 $mW/cm^2$. Polarization optics were then employed to separate the return beam and inject it into a polarization-maintaining fiber. Once in fiber this beam was combined with a fiber-coupled portion of the initial single-frequency laser output which served as a local oscillator.

The resulting self-heterodyne signal (see Fig. 2) was then observed via a fiber-coupled photodetector with a power bandwidth (3 dB) of 1 GHz, a conversion gain of 360 V/W, and a noise-equivalent power of 31 $pW/Hz^{1/2}$ (at 760 nm). This signal was amplified and split between a digitizer board for data acquisition and the phase-locking electronics. The 12-bit digitizer board coherently averaged 40 000 interferograms recorded in 100 μs segments at 3 GSamples/s (i.e., 200 ms of data acquisition). Due to the high data volumes, the throughput rate was 0.06 %, although this likely could be improved by more sophisticated data processing protocols. Spectral normalization was performed through the use of spectra which were recorded at a frequency detuned from the region of interest by 3 GHz.

Representative self-heterodyne absorption spectra of $^{39}$K can be found in Fig. 3. As the majority of the optical power is found at the optical carrier frequency ($\nu_0$), this tone acts as a pump in our pump-probe saturation spectroscopy measurements. Six distinct sets of sub-Doppler features are superimposed upon the Doppler profile. As expected, there is a large hole burning (HB) feature at the optical frequency, $\nu_0$, of the strong carrier tone. In addition, there are two sets of peaks present at $\nu_0 \pm 462$ MHz which arise from hyperfine pumping (HP) [24, 25] of the 4 $^2S_{1/2}$ ground state hyperfine levels through the upper state. Importantly, measurements of the frequencies of these HP peaks at $\nu_0 \pm 462$ MHz relative to $\nu_0$ provides for a multiplexed measurement of the lower-state hyperfine splitting without the need for an absolute frequency reference. In addition, the 4 $^2P_{1/2}$ hyperfine levels can be observed as a splitting of these features. Therefore, sub-Doppler spectroscopy with an optical frequency comb is capable of measuring both lower- and upper-state hyperfine splittings simultaneously with an uncertainty fundamentally limited by that of the radiofrequency comb generation signal and therefore traceable to the International System of Units (SI).

In addition to the above features we can also observe electromagnetically induced transparency (EIT) resonances [26-28] in the central HP peaks (see Fig. 4). These ultranarrow resonances have full-widths at half maximum of a few MHz making it possible to measure the ground state splitting with a precision at the tens-of-kHz level. We note that to first order the EIT resonances are independent of laser linewidth due to the common-mode nature of this self-heterodyne measurement [28]. The addition of magnetic-interference-shielding nickel-iron-cobalt foil was observed to reduce the width of these features by a further factor of approximately two through a reduction in Zeeman splitting due to stray magnetic fields. High-precision fits of the HP and EIT resonances repeated over several days in combination with a conservative estimate of potential systematic sources of uncertainty (e.g., dynamic Stark effect and Zeeman effect) yields values for the lower- and upper-state hyperfine splittings for the $D_1$ transition of $\Delta'' = 461.83(41)$ MHz and $\Delta' = 55.04(41)$ MHz, respectively with the corresponding combined standard uncertainties.

The sub-Doppler features so far described do not require counter-propagating beams as they are simply a result of strong optical pumping. Absolute transition frequencies of the hyperfine transitions [29] can be determined using the remaining three sets of saturation features which occur because of the counter-propagating beams (and are labelled with their corresponding assignments in Fig. 3). In this case, for a given optical carrier frequency, $\nu_0$, there exists a velocity class with a Doppler-shifted optical frequency of $\nu_0 = \nu_m + \Delta\nu_d$ where $\nu_m$ is the resting frequency of a given hyperfine transition and $\Delta\nu_d = -\nu_m(\boldsymbol{k}\cdot\boldsymbol{v_b})/(2\pi)$ is the Doppler shift, in which $\boldsymbol{k}$ is the laser wave vector and $\boldsymbol{v_b}$ is the velocity class vector. The saturation of this velocity class is then observed by the counter-propagating optical frequency comb with wave vector $\boldsymbol{k}$ of opposite sign at $\nu_{obs} = \nu_m - \nu_d = 2\nu_m - \nu_0$. As a result, in Fig. 3 we can observe sub-Doppler features resulting from transitions originating in each of the two hyperfine levels of the ground state as well as the cross-over resonances. We note that high-accuracy absolute frequencies could readily be measured by recording the beat frequency between the external-cavity diode laser and an absolute optical frequency reference. In addition, for high-resolution applications in chemical dynamics where the upper and lower state splittings are the desired quantities, absolute transition frequencies may not be required.

The comb-based approach we have demonstrated herein offers significant advantages over traditional single-frequency approaches. First, it is inherently multiplexed and allows for each of the hyperfine transitions to be recorded and quantified simultaneously. In addition, neither detailed *a priori* knowledge of the hyperfine levels nor high-resolution scanning of the pump laser frequency are required. Critically, the saturation features are observable as long as the carrier frequency is within the Doppler envelope of the absorption features. This is particularly important for systems with blended transitions where the component frequencies are not presently well known (e.g., congested spectra such as $CH_4$, the $NO_3$ radical, or floppy molecules such as $CH_5^+$). AWGs with sampling rates as high as 92 GSamples/s are commercially available, which when paired with commercial high-bandwidth phase modulators could further broaden the optical bandwidth to >60 GHz. Further bandwidth enhancements could be achieved by imprinting the present electro-optic-phase-modulator-based frequency comb upon a mode-locked laser frequency comb [13]. Extension of the present approach to the mid-infrared region could be achieved using difference frequency generation [30] approaches and should allow for rapid, multiplexed sub-Doppler spectra on a variety of important molecular species.

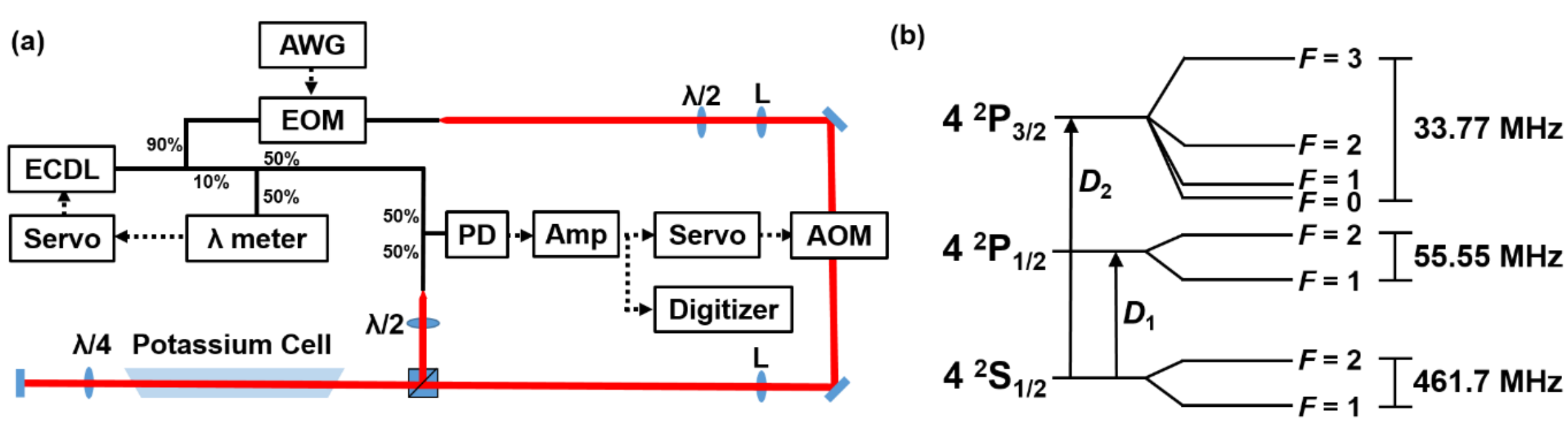

FIG. 1. (a) Schematic of the sub-Doppler direct frequency comb spectrometer. Fiber-coupled optical paths are shown in solid black, free-space optical paths in red, and electronic signals with dashed black arrows. The abbreviated elements are an external-cavity diode laser (ECDL), an arbitrary waveform generator (AWG), an electro-optic phase modulator (EOM), lenses (L), an acousto-optic modulator (AOM), a photodiode (PD), and radiofrequency amplifiers (Amp). The ECDL frequency is locked to a high-resolution wavelength meter by feeding back to the laser's current and piezoelectric transducer. The down-converted carrier frequency is phase-locked to an external radiofrequency reference, thus allowing for coherent self-heterodyne signal averaging. The AWG and all other microwave electronics were referenced to a 10 MHz signal from a commercial rubidium atomic clock. (b) Energy level diagram for $^{39}$K showing the $D_1$ and $D_2$ electronic transitions at 770.1 nm and 766.7 nm, respectively [29, 31].

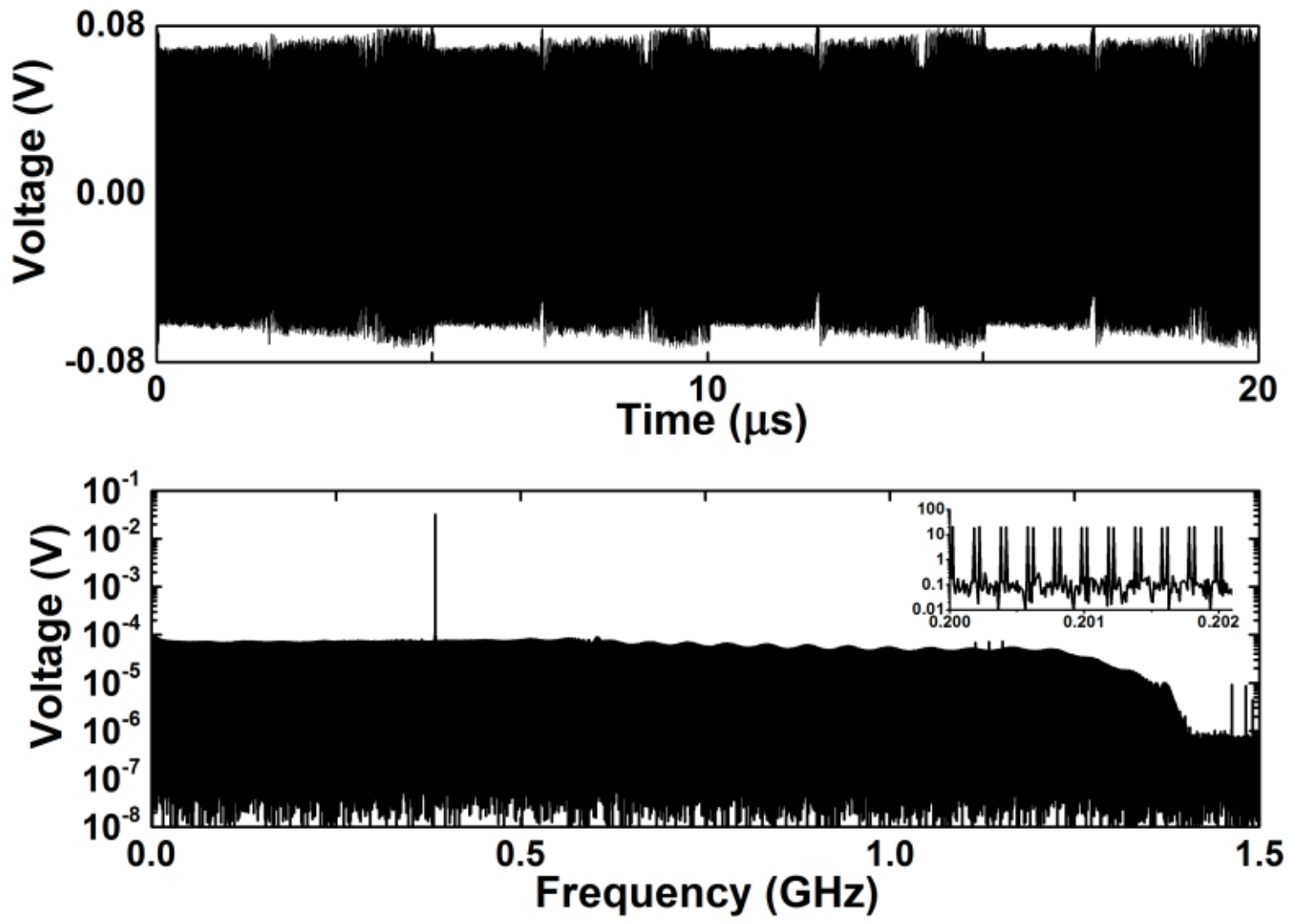

FIG. 2. (Upper panel) A portion of the 100-μs-long time domain self-heterodyne optical signal after 2,000 coherent averages. The interferogram repeats every 5 μs. (Lower panel) The corresponding frequency domain spectrum after a Fourier transform of the entire averaged optical heterodyne signal. The series of repeated time domain waveforms gives rise to an optical frequency comb. Ten thousand individual comb teeth are present with a spacing of 0.2 MHz (i.e., the inverse of the waveform duration). The lower optical frequency teeth are wrapped about DC. The carrier (i.e., acousto-optic modulator frequency) can be seen at 0.38422 GHz.

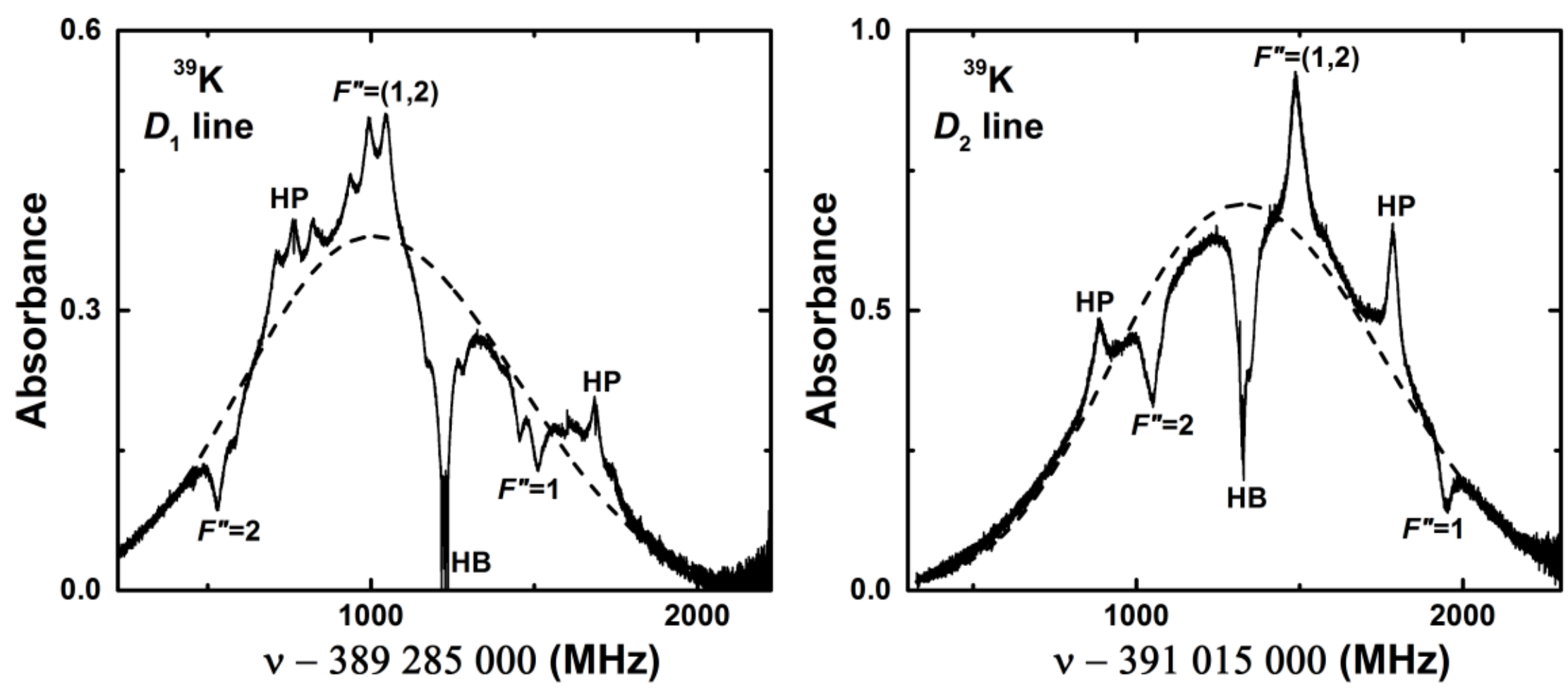

FIG. 3. Sub-Doppler spectra of the $^{39}$K $D_1$ and $D_2$ lines. Two thousand 100 μs long time domain optical self-heterodyne signals were coherently averaged, followed by subsequent averaging of 40 and 22 normalized spectra, respectively. Counter-propagating features are labelled with their assignments, while co-propagating features are labelled as hole burning (HB) and hyperfine pumping (HP). For reference, the calculated Doppler-broadened envelope is shown with the dashed line. Electromagnetically induced transparency (EIT) can be seen in the central HP features of the $D_1$ line (see also Fig. 4). Third-order and linear baselines were subtracted from the $D_1$ and $D_2$ spectra, respectively.

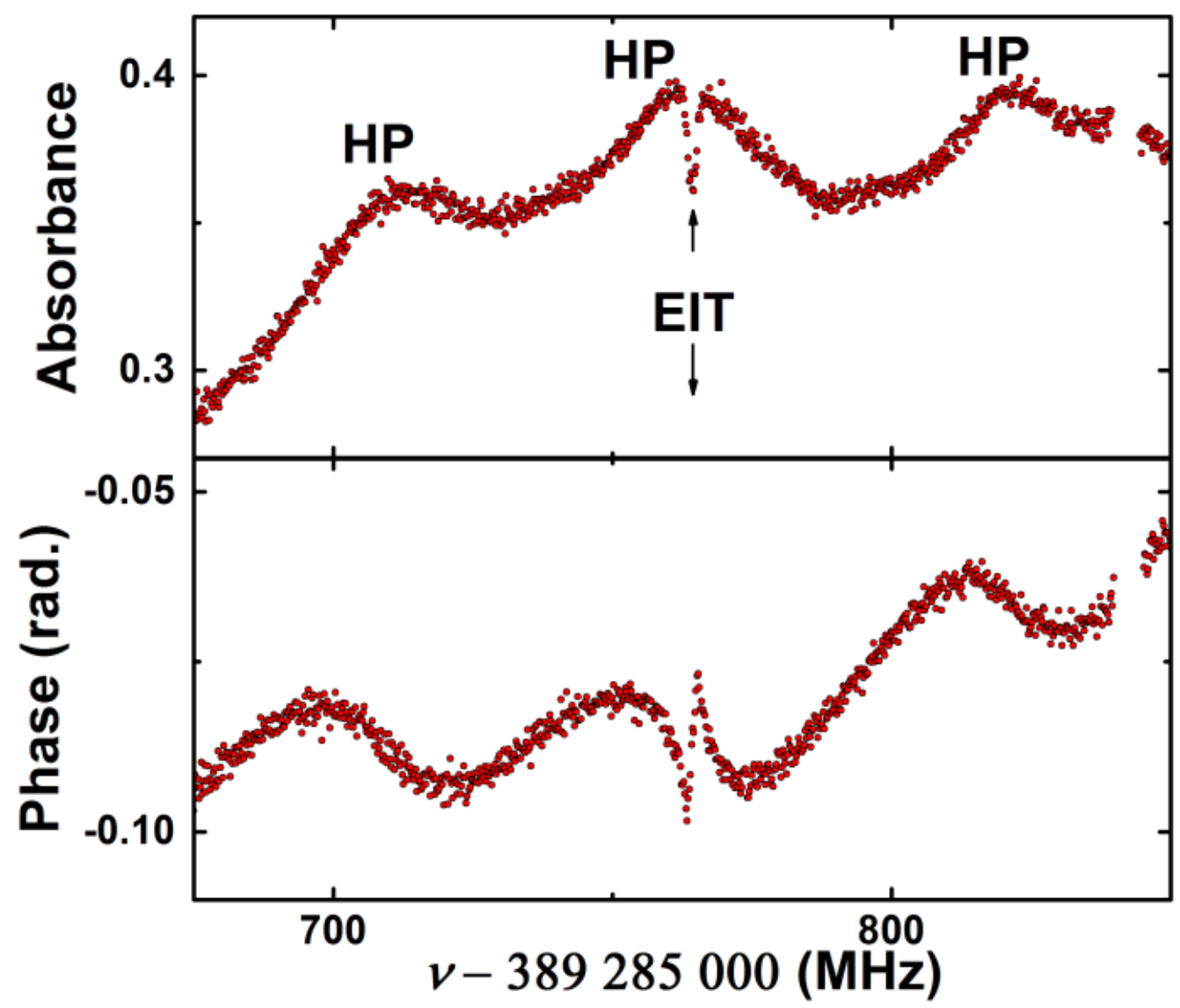

FIG. 4. A portion of the absorption and phase spectra for the $^{39}$K $D_1$ line. The absorption spectrum is a subset of the spectrum shown in Fig. 3. Two thousand 100 μs long time domain self-heterodyne optical signals were coherently averaged, followed by subsequent averaging of 40 spectra. In addition to the hyperfine pumping (HP) peaks there is clear evidence of electromagnetically induced transparency (EIT). A small number of points near 389 285 850 MHz were removed from the shown figure as these points occur near to DC in the optical heterodyne signal. A third-order linear baseline was subtracted from the absorption spectrum.

## Acknowledgements

Funding was provided by the National Institute of Standards and Technology (NIST) Greenhouse Gas Measurement and Climate Research Program. We also thank K. C. Cossel, J. R. Lawall, N. Newbury, J. N. Tan, and J. Unguris of NIST for loaned equipment.

**Supplementary Material**

The probe comb transmitted through the potassium vapor cell was combined on a fast photoreceiver with the original continuous-wave laser that seeded the chirped-waveform-driven phase modulator comb (see Fig. 1a of the main text). When the comb was tuned to be on resonance with a potassium absorption feature, this self-heterodyne signal, normalized to the self-heterodyne signal measured with the comb off resonance, provided both amplitude and phase spectra of the potassium vapor. The amplitude and phase spectra were modeled by a complex absorbance $\tilde{A}$, where the amplitude spectrum $|\tilde{A}| = \sqrt{\tilde{A} \cdot \tilde{A}^*}$ and the phase spectrum $\phi = \tan^{-1}\left(\mathrm{Im}(\tilde{A})/\mathrm{Re}(\tilde{A})\right)$.

The multiplexed pump-probe comb spectra shown in Fig. 3 of the main text include both Doppler-broadened and sub-Doppler absorption features. As a worked example, this Supplemental Material describes in detail the origin of each peak and dip observed for the $^{39}$K $D_1$ line. For reference, absolute transition and cross-over frequencies $\nu_m$ from the literature are listed in Table S1 [1]. The complex absorbance $\tilde{A}$ as a function of frequency $\nu$ for the $D_1$ line is:

$$\begin{aligned}\tilde{A}(\nu) = &\sum_{m=1}^{M} \tilde{a}_m \tilde{g}_V(\nu; \nu_m, \gamma_D, \gamma_L) \\ &+ \tilde{b}\tilde{g}_L(\nu; \nu_0, \gamma_{HB}) \pm \tilde{b}_s \tilde{g}_L(\nu; \nu_0 + \Delta', \gamma_{HB}) \\ &+ \sum_{n=-N+1}^{N-1} \tilde{c}_n \tilde{g}_L(\nu; \nu_0 \pm \Delta'' + n\Delta', \gamma_{HP}) \\ &+ \sum_{m=1}^{M} \tilde{d}_m \tilde{g}_L(\nu; 2\nu_m - \nu_0, \gamma_s)\end{aligned} \qquad \text{(S1).}$$

The first term on the right-hand side of the above Eq. S1 is a sum over all Doppler-broadened hyperfine transitions $M$. The four unique transitions (designated labels $m$ = 1, 3, 7 and 9 in Table S1) were each modeled with a complex Voigt function $\tilde{g}_V$ [2] with inhomogeneous half-width at half maximum (HWHM) of $\gamma_D$ = 385 MHz and a homogeneous HWHM of $\gamma_L \approx 0$. The relative complex amplitudes $\tilde{a}_m$ of the four hyperfine transitions were fixed to relative values of 1.0:1.0:1.0:0.2 [3] and then scaled by a single floated coefficient.

**Table S1.** Transition and cross-over frequencies of the $^{39}$K $D_1$ lines [1].
An offset of 389 285 000 MHz was subtracted from $\nu_m$.

| *m* | *F″* | *F′* | $\nu_m$ (MHz) |
|---|---|---|---|
| 1 | 2 | 1 | 850.838 |
| 2 | 2 | (1, 2) | 878.612 |
| 3 | 2 | 2 | 906.386 |
| 4 | (1, 2) | 1 | 1081.706 |
| 5 | (1, 2) | (1, 2) | 1109.481 |
| 6 | (1, 2) | 2 | 1137.255 |
| 7 | 1 | 1 | 1312.574 |
| 8 | 1 | (1, 2) | 1340.349 |
| 9 | 1 | 2 | 1368.124 |

The second term on the right-hand side of Eq. S1 is the hole burned into the Doppler-broadened profile by the strong pump comb tooth at an absolute optical frequency of $\nu_0 = \nu_{wm} + f_{AOM} + f_0$, where $\nu_{wm}$ is the optical frequency measured by the wavelength meter, $f_{AOM} = 384.22$ MHz (see main text), and $f_x$ is a floated wavelength meter offset ($f_x \approx 190$ MHz). The line shape chosen for the hole-burning (HB) feature was a complex Lorentzian function $\tilde{g}_L$ with HWHM $\gamma_{HB}$ and an amplitude term $\tilde{b}$. Both $\gamma_{HB}$ and $\tilde{b}$ were both floated during fitting. The hole burning peak has two satellite peaks at $\nu_0 \pm \Delta'$ due to saturation and four-wave mixing in a *V*-type three level subsystem [4]. Their complex amplitudes were both fit to the same value of $\tilde{b}_s$.

The third term on the right-hand side of Eq. S1 is the hyperfine pumping (HP) term which describes population transfer between the lower-state hyperfine levels via strong velocity-selective optical pumping [4-6]. These HP sub-Doppler features are labeled in Fig. 3 of the main text, and appear clustered at $\nu_0 \pm \Delta''$, where $\Delta'' = 462$ MHz is the lower-state (4 $^2S_{1/2}$) hyperfine splitting (see Fig. 1b of the main text). These two clusters at $\nu_0 \pm \Delta''$ are further split by the upper-state (4 $^2P_{1/2}$) hyperfine splitting $\Delta' = 56$ MHz. There are $n$ number of peaks in each HP cluster, where $n = -N+1, -N+2, \ldots N-1$ and $N = 2$ is the number of hyperfine levels in the upper state. Therefore, for the $^{39}$K $D_1$ lines $n = -1, 0$ or $1$. The individual features were again modeled by complex Lorentzian functions $\tilde{g}_L$ with floated coefficients $\tilde{c}_m$ and a common HWHM $\gamma_{HB}$. For simplicity, we set $\gamma_{HP} = \gamma_s$, the HWHM of the sub-Doppler features described by the fourth term in Eq. S1. When modeling the $D_2$ lines there are several upper-state splittings that must be considered, and therefore additional terms similar to $n\Delta'$ shown in Eq. S1 must also be included.

For the special case of $n = 0$ in both HP features of the $D_1$ line, electromagnetically induced transparency (EIT) is observed. Therefore, a modified complex Lorentzian function was used which is related to the first-order atomic susceptibility ($\chi^{(1)}$) of a $\Lambda$-type three-level subsystem under strong optical pumping [7-9]. In the $\Lambda$-type subsystem defined by two lower-state hyperfine levels and one upper-state hyperfine level, it is reasonable to make the steady-state approximation that all of the potassium atoms are in the two lower-state hyperfine levels ($C_1 + C_2 = 1$ and $C_3 = 0$ in the nomenclature of Ref. 9, where $C_x$ are the coefficients for each level *x*). However, as long as the probe power is low ($\Omega_p << \Omega_c$, where $\Omega$ is the Rabi frequency for either the probe (*p*) or the cycling ($c$) transitions), the EIT lineshape function reduces to that routinely found in the literature [7-9].

The fourth term describes sub-Doppler features associated with the absolute transition and cross-over frequencies $\nu_m$ which are essentially the strong-pump-weak-probe manifestations of the "double hole-burning effects" discussed by Siegman [10]. Before floating, an approximate value of $\gamma_s$ was calculated via the standard saturated absorption HWHM formula $\gamma_s = \gamma\left(1 + \sqrt{1 + I/I_s}\right)$ given by Demtröder [11] where $\gamma = 3.0$ MHz is the natural line width and $I_s \approx 3$ mW/cm$^2$ is an estimate of the saturation intensity [12, 13]. Again, a complex Lorentzian function was chosen for modeling, and all coefficients $\tilde{d}_m$ where floated during fitting. Eq. S1 also highlights an important feature of this original experimental approach: the hyperfine pumping sub-Doppler peaks in the third term of Eq. S1 provide a direct measurement of $\Delta''$ and $\Delta'$ simultaneously and without the need to independently measure each absolute transition frequency.

Over a period of several days, numerous (>100) sub-Doppler spectra of the $D_1$ line were recorded as described in the main text. For each spectrum, an average value for $\Delta''$ was measured using both of the EIT features at $\nu_0 \pm \Delta''$. Additionally, an average value for $\Delta'$ was measured using all available sub-Doppler features (HP, saturation, and four-wave mixing). The weighted averages of all measured values of $\Delta''$ and $\Delta'$ are reported in the main text. The standard deviation of the weighted average is reported in Table S2 as the uncertainty in the experiment fit.

The largest two potential sources of systematic uncertainty arise from stray DC magnetic fields (Zeeman effect) and from strong optical pumping (dynamic Stark effect). Magnetic shielding reduced stray magnetic fields to less than 200 mG (limited by the resolution of the available Gaussmeter), leading to the estimated uncertainty in Table S2. The magnitude of the dynamic Stark effect was estimated using twice the largest observed value of the Rabi frequency for the optical pump ($\Omega_c$) at $\nu_0$, a direct measure of which was provided by the sub-natural linewidth EIT. The photon recoil effect is small compared to our estimates of the Zeeman and dynamic Stark effects, and therefore was not included in the overall uncertainty budget, and the laser polarization was considered to be purely linear with perpendicularly oriented counter-propagating orientations. The uncertainty budget estimates reported below are consistent with the values reported in Ref. 1 and 14 given the differences in experimental conditions.

**Table S2.** Uncertainty budget and combined standard uncertainty for reported values of $\Delta''$ and $\Delta'$.

| Source | Uncertainty (1σ) |
|---|---|
| experimental fit ($\Delta''$) | 200 kHz |
| experimental fit ($\Delta'$) | 190 kHz |
| Zeeman effect | 280 kHz |
| dynamic Stark effect | 230 kHz |
| sum in quadrature ($\Delta''$) | 410 kHz |
| sum in quadrature ($\Delta'$) | 410 kHz |